\documentstyle[12pt]{article} 
\begin{document}
\renewcommand{\theequation}{\arabic{section}.\arabic{equation}}
\newcommand{\beq}{\begin{equation}}
\newcommand{\eeq}{\end{equation}}
\newcommand{\beqn}{\begin{eqnarray}}
\newcommand{\eeqn}{\end{eqnarray}}
\newcommand{\slp}{\raise.15ex\hbox{$/$}\kern-.57em\hbox{$\partial
$}}
\newcommand{\slA}{\raise.15ex\hbox{$/$}\kern-.57em\hbox{$A$}}
\newcommand{\lnA}{\raise.15ex\hbox{$/$}\kern-.57em\hbox{$A$}}
\newcommand{\slB}{\raise.15ex\hbox{$/$}\kern-.57em\hbox{$B$}}
\newcommand{\bP}{\bar{\Psi}}
\newcommand{\bC}{\bar{\chi}}
 \newcommand{\hs}{\hspace*{0.6cm}}

\title{Non-local Thirring model at finite temperature}
\author{ 
M.V.Man\'{\i}as$^{a,b}$,C.M.Na\'on$^{a,b}$, and M.L.Trobo$^{a,b}$}
\date{December 1997}
\maketitle

\def\thepage{\protect\raisebox{0ex}{\ } La Plata 97-25}
\thispagestyle{headings}
\markright{\thepage}

\begin{abstract}

 We extend a recently proposed non-local and non-covariant version of the 
Thirring model to the finite-temperature case.\\
\hs We obtain a completely bosonized expression for the partition function, 
describing the thermodynamics of the collective modes which are the underlying 
excitations of this system. From this result we derive closed formulae for the 
free-energy, specific-heat, two-point correlation functions and momentum 
distribution, as functionals of 
electron-electron coupling potentials. 
\end{abstract}
 
\vspace{3cm}
Pacs: \\ 
\hspace*{1,7 cm} 11.10.-z \\
\hspace*{1,7 cm} 11.15.-q

\noindent --------------------------------

\noindent $^a$ {\footnotesize Depto. de F\'\i sica.  Universidad
Nacional de La Plata.  CC 67, 1900 La Plata, Argentina.\\
E-mail: naon@venus.fisica.unlp.edu.ar}

\noindent $^b$ {\footnotesize Consejo Nacional de Investigaciones
Cient\'\i ficas y T\'ecnicas, Argentina.}

\newpage
\pagenumbering{arabic}

\section{Introduction}

\hs In recent years it has been witnessed a renewed interest in the study of low-
dimensional field theories. In particular, research on the one-dimensional (1d) 
fermionic gas has been very active, mainly due to the actual 
fabrication of
the so called quantum wires \cite{Voit}. One of the most interesting aspects
of these systems is the possibility of having a deviation from the usual
Fermi-liquid behavior. This phenomenon was systematically examined by Haldane 
\cite{Haldane} who coined the term Luttinger-liquid behavior to name this
new physical situation in which the Fermi surface disappears and the spectrum
contains only collective modes. Perhaps the simplest theoretical framework 
that presents this feature is the Tomonaga-Luttinger (TL) model \cite{TL}, 
a many-body system of right and left-moving particles interacting through their charge densities. In a recent series of papers \cite{NLT} an alternative, field-theoretical approach was developed to consider this problem. In these works a non-local and non-covariant version of the Thirring model was introduced, in which the fermionic densities and currents are coupled through
bilocal, distance-dependent potentials. This non-local Thirring model (NLT) 
contains the TL model as a particular case. Although it constitutes an elegant framework to analyze the 1d many-body problem, one seriuos limitation appears if one tries to make contact with quantum wires phenomenology. Indeed, one has to recall that the NLT has been formulated at zero temperature. This means, of course, that as it stands it cannot be used to study the Luttinger 
liquid thermodynamics. The main purpose of this paper is to fill this gap. 
To this end we employ the well-known imaginary time formalism \cite{ber} \cite{mat} 
in order to obtain the finite-temperature version of the NLT model. In Section 2
we verify that the manipulations used to write the vacuum functional at $T = 0$ 
in terms of a fermionic determinant, also work for the NLT at an arbitrary 
equilibrium temperature. Using the so called decoupling technique at finite 
temperature \cite{mnt} we obtain the partition function which describes the 
thermodynamics of the effective bosonic degrees of freedom (charge density and 
spin-density waves). In Section 3 we derive expressions for the Helmholtz free energy,
the energy and the specific heat as functionals of the forward-scattering potentials. In Section 4 we compute
the two-point fermionic correlation function and using this result, in Section 5 we obtain the formula for the 
momentum distribution. One of the most interesting aspects of these formulae is that
being functionals of the potentials they could be employed as starting points 
in order to perform quantitative tests of different electron-electron couplings.
Of course, these studies could be analytical or numerical, depending on the 
specific potentials to be considered. Finally, in Section 6, we summarize the 
main points of our investigation.                                                          

\section{The partition function}
\setcounter{equation}{0}
\hs In this section we study the two-dimensional non-local Thirring model 
\cite{NLT}
at finite temperature using the imaginary time formalism developed by
Bernard \cite{ber} and Matsubara \cite{mat}. Our starting point is the Euclidean
action given by

\beq
 S = \int_{\beta} d^2x~ \bP i \slp \Psi  - \frac{g^2}{2} \int_{\beta} d^2x d^2y ~
[J_{\mu}^a(x)V_{(\mu)}^{ab}(x-y) J_{\mu}^b(y) ]
\label{1}
\eeq
where $\int_{\beta} d^2x~$ means $\int_0^{\beta}dx^0 \int dx^1$ and 
$\beta=\frac{1}{k_{B}T}$  with $k_{B}$ the Boltzman's constant and 
$T$ the temperature. 
The fermionic current is represented by
$J_{\mu}^a = \bP \gamma_{\mu}\lambda^a \Psi$ 
with $\lambda^0 = \frac{1}{2}I$, $\lambda^j = t^j$, $t^j$ being the
SU(N) generators normalized according to $tr(t^it^j) = \delta^{ij}/2$. In
order to make direct contact with a system of 1d electrons, we introduce the
Fermi velocity $v_{F}$ by defining the $\gamma_{\mu}$ matrices as

\beq
\gamma_{0} = \left( \begin{array}{cc}
0 \;\; 1 \\ 
1 \;\; 0 \\
\end{array} \right)
\label{2}
\eeq

\beq
\gamma_1 =\left( \begin{array}{cc}
0 \;\; iv_{F}\\
-iv_{F} \;\; 0 \\
\end{array} \right)
\label{3}
\eeq
Of course, for $v_{F}=1$ one reobtains the set of matrices usually employed 
in $1+1$ QFT's.
The functions $V_{(\mu)}^{ab}(x-y)$ are $N^2\times N^2$ matrices whose elements are the potentials describing electron-electron forward-scattering interactions. Let us anticipate that, although the procedure that follows
works for arbitrary $N$, we shall be specially interested in the particular case 
$N = 2$, which is the natural choice to consider spin-$\frac{1}{2}$ particles in 
this non-relativistic framework. To avoid confusion let us also note that no sum 
over repeated indices will be implied when a subindex $(\mu)$ is involved. 
In other words, the interaction between currents appearing in the action above 
reads

\beq
J_{\mu}^a(x)V_{(\mu)}^{ab}(x-y) J_{\mu}^b(y) =
J_{0}^a(x)V_{(0)}^{ab}(x-y) J_{0}^b(y) +
J_{1}^a(x)V_{(1)}^{ab}(x-y) J_{1}^b(y) 
\label{4}
\eeq

We shall consider the vacuum functional

\beq
Z= N N_{F}(\beta) \int_{antiper} D\bP~ D\Psi e^{-S} 
\label{5}
\eeq
where $N$ is an infinite $\beta$-independent constant while $N_{F}(\beta)$ is 
a $\beta$-dependent infinite factor that will be determined later on in this Section.
The functional integral in (\ref{5}) must be extended over the paths with 
antiperiodicity conditions in the Euclidean time variable $x^0$:

\beqn
\Psi (x^0+\beta, x^1) & = & - \; \Psi(x^0, x^1) \nonumber\\
\bP~ (x^0+\beta, x^1) & = & - \; \bP~(x^0, x^1) 
\label{6}
\eeqn

 Exactly as one does in the usual (local and covariant) Thirring model, the 
fermionic quartic interaction can be eliminated by introducing auxiliary vector 
fields. In this way one can express the partition function in terms of a 
fermionic determinant. This, in turn, allows to implement the path-integral 
approach to non-local bosonization \cite{NLT}, which we want to extend to the 
$T\neq{0}$ case. To follow this procedure it is convenient to split S in the form

\beq
S = S_0 + S_{int}
\label{7}
\eeq
\noindent where
\beq     
S_0 = \int_{\beta} d^2x~ \bP i\slp\Psi,
\label{8}
\eeq
\noindent and
\beq
S_{int} = -\frac{g^2}{2} \int_{\beta} d^2x~ J_{\mu}^a K_{\mu}^a.
\label{9}
\eeq
In this last expression $K_{\mu}^a$ is a new current defined as

\beq
K_{\mu}^a(x) = \int_{\beta} d^2y~ V_{(\mu)}^{ab}(x,y)J_{\mu}^b(y).
\label{10}
\eeq

The partition function can now be written as

\beq
Z = N N_{F}(\beta) \int_{_{\rm antiper}} D\bP D\Psi ~\exp \{ -\int_{\beta} d^2x~\bP 
i\slp\Psi  ~ + \frac{g^2}{2} \int_{\beta} d^2x~ J_{\mu}^{a} K_{\mu}^{a} \}
\label{11}
\eeq

Now, we shall introduce a vector field $\tilde{A}_{\mu}$ in the form:
\beq
\int D\tilde{A}_{\mu}^{a} \delta(\tilde{A_{\mu}^{a}} - K_{\mu}^{a})
\exp \{ \frac{g^2}{2} \int_{\beta} d^2x \tilde{A}_{\mu}^{a}J_{\mu}^{a} \}=
\exp \{ \frac{g^2}{2} \int_{\beta} d^2x K_{\mu}^{a}J_{\mu}^{a} \}
\label{12}
\eeq
We represent the delta functional using a $\tilde{B}_{\mu}$-field as follows
\beq
\delta (\tilde{A}_{\mu}^{a}- K_{\mu}^{a}) = \int D\tilde{B}_{\mu}^{a} exp \{- \int_{\beta} 
d^2x \tilde{B}_{\mu}^{a}(\tilde{A}_{\mu}^{a} - K_{\mu}^{a}) \}
\label{13}
\eeq

We have to impose periodicity conditions for the bosonic $\tilde{A}_{\mu}$ 
and $\tilde{B}_{\mu}$-fields over the range $[0,\beta]$. Using now (\ref{12}) 
and (\ref{13}), the fermionic piece of the action can be written as

\beq
S_{0} - \frac{g^2}{2}\int_{\beta} d^2x (\tilde{A}_{\mu}^a
J_{\mu}^a + \frac{2}{g^2} \tilde{B}_{\mu}^aK_{\mu}^a)  =
\int_{\beta} d^2x \bP~( i\slp -\frac{g^2}{2} \gamma_{\mu} (\tilde{A}_{\mu}^a +
\bar{B}_{\mu}^a)) ~\Psi
\label{14}
\eeq      
where we have defined 
\beq
\bar{B}_{\mu}^a(x) =\frac{2}{g^2} \int_{\beta} d^2y~ V_{(\mu)}^{ab}(y,x)\tilde{B}_{\mu}^b(y).
\label{15}
\eeq
For later convenience we shall invert (\ref{15}) in the form

\beq
\tilde{B}_{\mu}^a(x) = \frac{g^2}{2}\int_{\beta} d^2y~ b_{(\mu)}^{ab}(y,x) \bar{B}_{\mu}^b(y),
\label{16}
\eeq
with $b_{(\mu)}^{ab}(y,x)$ satisfying  

\beq
\int d^2y~ b_{(\mu)}^{ab}(y,x) V_{(\mu)}^{bc}(z,y) = \delta^{ac}\delta^2 (x-z).
\label{17}
\eeq
 At this point we make the change

\beq
\frac{g}{2}(\tilde{A}_{\mu}^a +\bar{B}_{\mu}^a) = A_{\mu}^a
\label{18}
\eeq       
    
\beq
\frac{g}{2}(\tilde{A}_{\mu}^a - \bar{B}_{\mu}^a) = B_{\mu}^a,
\label{19}
\eeq
which allows us to write 
\beqn
Z & = & N_1 N_{F}(\beta) ~\int_{_{\rm periodic}} DA_{\mu}^a DB_{\mu}^b~ det_{\beta}(i~\slp - 
g~\lnA) \times \nonumber\\
& \times & exp \{ \frac{-1}{2}~ \int_{\beta} d^2x d^2y
[ b_{(\mu)}^{ab}(y,x)A_{\mu}^{a}(x) A_{\mu}^{b}(y) - \nonumber \\ 
& - & b_{(\mu)}^{ab}(x,y)B_{\mu}^{a}(x) B_{\mu}^{b}(y)] \}, 
\label{20}
\eeqn
where we have used the fact that the functions $V_{(\mu)}^{ab}$ and 
$b_{(\mu)}^{ab}$ are symmetric in coordinates.
Note that the Jacobian associated with the change $(\tilde{A},\tilde{B}) 
\rightarrow(A,B)$, does not depend neither on fields nor on temperature 
and then it can be absorbed in the normalization
constant $(N \rightarrow N_1)$. 

We have been able to express $Z$ in terms of a temperature-dependent fermionic 
determinant. This fact will enable us to apply the non-local bosonization 
scheme, first developed for $T = 0$ \cite{NLT}, to the present 
finite-temperature case.\\
Note that, as a consequence of the change of bosonic variables
(eqs.(\ref{18}) and (\ref{19})), the effect of the non-local interaction 
has been completely transfered
to the purely bosonic piece of the action, $S[A,B]$. On the other hand we see
that the field $B_{\mu}$ is completely decoupled from both the $A_{\mu}$-field 
and the fermion field. This clearly indicates that its contribution should be 
factorized and absorbed in the normalization constant.
However, the issue is more subtle since for repulsive interactions $B_{\mu}$ 
corresponds to a negative-metric state. This is not a peculiar feature of 
neither non-local nor finite-temperature theories. Indeed, the appearance of
negative-metric states was already stressed by Klaiber in his seminal
work on the usual Thirring model \cite{Klaiber}. In order to have a well defined 
Hilbert space, Klaiber had to disregard these fields. Following the same 
prescription in this new context, we are naturally 
led to include the decoupled $B_{\mu}$-integral in $N_{F}(\beta)$. Indeed, as it 
is habitual in finite-temperature studies, $N_{F}(\beta)$ contains all the 
$\beta$-dependent infinite contributions to the partition function. Since the 
integral over ghost-fields is also a $\beta$-dependent infinite factor 
(otherwise one would not reproduce the well-known result for the $T\neq0$ local 
Thirring model (See eq.(\ref{62}))), it is natural to include it in $N_{F}(\beta)$.
The partition function (\ref{20}) then reads

\beq
Z = N_{F}(\beta)~N_{1} \int_{_{\rm periodic}} DA_{\mu}^a 
det(i \slp - g \lnA) e^{-S[A_{\mu}]},
\label{21}
\eeq

\noindent where $S[A]$ is the $A_{\mu}$-dependent part of (\ref{20}).
From now on we shall take the fermion fields in the 
fundamental representation of the maximal abelian subgroup of U(2). In the 
many-body language this corresponds to a system of spin-~$\frac{1}{2}$ fermions 
in which spin-flipping processes are forbidden.\\
Now, the potential matrices are diagonal whose elements can be written in terms 
of S\'olyom's "g-ology" \cite{So} as

\begin{eqnarray}
V_{(0)}^{00}&=&\frac{1}{4}(g_{4 \parallel}+ g_{4 \perp} + g_{2 \parallel} 
+ g_{2 \perp}),\nonumber\\
V_{(0)}^{11}&=&\frac{1}{4}(g_{4 \parallel}- g_{4 \perp} + g_{2 \parallel} 
- g_{2 \perp}),\nonumber\\
V_{(1)}^{00}&=&\frac{1}{4}(-g_{4 \parallel}- g_{4 \perp} + g_{2 \parallel} 
+ g_{2 \perp}),\nonumber\\
V_{(1)}^{11}&=&\frac{1}{4}(-g_{4 \parallel} + g_{4 \perp} + g_{2 \parallel} 
- g_{2 \perp}),
\label{22}
\end{eqnarray}
where $g_2$ and $g_4$ are associated to scattering diagrams involving two or 
just one electronic species (left or right moving), respectively. On the other 
hand, the subscripts $\parallel$ and $\perp$ denote those processes in which 
incident fermions have parallel or anti-parallel spins.
Let us also recall for later convenience that the Tomonaga-Luttinger 
model, in which only charge-density fluctuations are considered, corresponds 
to $V_{(1)}^{00} = V_{(1)}^{11} =0$.\\
At this stage we can decouple the vector field $A_{\mu}$
from the fermion fields in the fermionic determinant contained in (\ref{21}). 
The only new feature in this otherwise standard step, is given by the modified 
$\gamma_1$ matrix of equation (\ref{3}). Fortunately this fact does not pose any 
substantial problem. Indeed, writing the components of $A_{\mu}$ in terms of two 
scalar fields $\phi$ and $\eta$ in the form

\beq
A_{0}(x) = v_{F}\partial_{1}\phi(x) - \partial_{0}\eta(x) 
\label{23}
\eeq

\beq
A_{1}(x) =\frac{-1}{ v_{F}}\partial_{0}\phi(x) - \partial_{1}\eta(x)
\label{24}
\eeq
with $\phi = \phi^{i} \lambda^{i}$, $\eta = \eta^{i} \lambda^{i}$, $i=0,1$,
it is easy to verify that the desired decoupling is achieved by the same chiral 
transformation in the fermionic variables that one would perform in the $v_F = 1$ 
case:

\beq
\Psi(x) = e^{g [\gamma_5 \phi(x) + i \eta(x)]} \chi(x)
\label{25}
\eeq

\beq
\bP(x) = \bar\chi(x) e^{g [\gamma_5 \phi(x) - i\eta(x)]}
\label{26}
\eeq
where $\gamma_5$ is the usual chiral matrix, i.e. the one that is obtained in 
the case $v_F = 1$.
Thus we get
\beq
det_{\beta} (i \slp - g \lnA) = 
J_F[\phi,\eta] det_{\beta} i \slp
\label{27}
\eeq
\noindent and the Jacobian of the fermionic transformation is given by
\cite{rd}

\beq
ln J_F[\phi,\eta] = \frac{-g^2}{2\pi} \int_{\beta} d^2x~[\frac{1}{v_{F}}
(\partial_{0}\phi)^2 + v_{F}(\partial_{1}\phi)^2] 
\label{28}
\eeq

Concerning the bosonic change of variables given by (\ref{23}) and (\ref{24}), 
one has to consider 
another temperature-dependent Jacobian, 
to be included in the path-integral measure as

\beq
DA_{\mu} = det_{\beta}(-{\Box})~D\phi^a ~D\eta^a
\label{29}
\eeq
where ${\Box} =\frac{1}{v_F} \partial_{0}^2 + v_{F} \partial_{1}^2$.\\
Inserting (\ref{27}), (\ref{28}), and (\ref{29}) in (\ref{21})
the result is

\beq
Z =N_{1} N_{F}(\beta)  det_{\beta}^2(i \slp) det_{\beta}^2({-\Box})
\int \prod_{a=0,1}D\phi^a D\eta^a \exp-S_{eff}^{aa}[\phi, \eta]
\label{30}
\eeq
where 
\beqn
S_{eff}^{aa}[\phi, \eta] & = & \frac{\lambda}{2} \int_{\beta} d^2x~ [\frac{1}{v_{F}}
 (\partial_{0}\phi^a)^2 + v_{F}(\partial_{1}\phi^a)^2] + \nonumber\\
 & + & \frac{1}{2}\int_{\beta} d^2x d^2y ~[v_{F}^2~\partial_1 \phi^a (x) 
 b_{(0)}^{aa}(x,y) \partial_1 \phi^a (y) + \nonumber \\
& + & \frac{1}{v_{F}^2}  
 \partial_0 \phi^a (x)b_{(1)}^{aa}(x,y) \partial_0 \phi^a (y) + 
 \partial_0 \eta^a b_{(0)}^{aa}(x,y)\partial_{0} \eta^a (y) + \nonumber\\
 & + & \partial_{1} \eta^a(x) b_{(1)}^{aa}(x,y)\partial_1 \eta^a (y) + 
  2 v_{F} \partial_{0} \eta^a(x) b_{(0)}^{aa}(x,y)\partial_{1} \phi^a(y) -
\nonumber\\ 
&-& \frac{2}{v_{F}} \partial_{1} \eta^a(x) b_{(1)}^{aa}(x,y) \partial_{0} 
 \phi^a (y)]
 \label{31}
 \eeqn
and $\lambda = \frac{g^2}{\pi}$.\\
In order to evaluate $ln Z$ we follow the pioneering work of
Bernard \cite{ber} and expand the bosonic fields $\phi(x^0,x^1)$ and 
$\eta(x^0,x^1)$, which
are periodic in the interval $0\leq x \leq \beta$, in Fourier series:

\beq
\phi^a(x^0,x^1)=\frac{1}{\beta}\sum_{n=-\infty}^{\infty}~\int~\frac{dk_1}
{2\pi}~e^{ik_1 x^1} e^{i\omega_{n}x^0}~\tilde{ \phi_{n}^a}(k_1)
\label{32}
\eeq

\beq
\eta^a(x^0,x^1)=\frac{1}{\beta}\sum_{n=-\infty}^{\infty} \int\frac{dk_1}
{2\pi}e^{ik_1 x^1} e^{i\omega_{n}x^0} \tilde{\eta_{n}^a}(k_1)
\label{33}
\eeq
where

\beq
\tilde{\phi_{n}^a}(k_1)=\int dx^1 \int_{0}^{\beta} dx^0 e^{-ik_1 x^1}e^{-i\omega_{n}x^0}
\phi^a(x^0,x^1)
\label{34}
\eeq

\beq
\tilde{ \eta_{n}^a}(k_1)=\int dx^1 \int_{0}^{\beta} dx^0 e^{-ik_1 x^1}e^{-i\omega_{n}x^0}
\eta^a(x^0,x^1)
\label{35}
\eeq
and the Matsubara frequencies are given by

\beq
\omega_{n}=\frac{2n \pi}{\beta}.
\label{36}
\eeq
In analogous way one can expand the inverse potentials as

\beq
b_{(\mu)}^{ab}(x^1,y^1,x^0,y^0) = \frac{1}{\beta}\sum_{n=-\infty}^{\infty}
\int\frac{dk'_1}{2\pi}
e^{ik'_1(x^1- y^1)}e^{i\omega_{n'}(x^0 -y^0)} \tilde {b}_{(\mu)n'}^{ab}(k'_1).
\label{37}
\eeq

In writing the above expressions we have used discrete and continuous delta 
functions defined by
\beq
\int_0^{\beta} dx^0 e^{i(\omega_{n} - \omega_{n'})x^0} = \beta \delta_{n,n'}
\label{38}
\eeq
\beq
\int \frac{dx^1}{2 \pi} e^{i(k_1 - k'_1)x^1} = \delta( k_1 - k'_1).
\label{39}
\eeq
Equation (\ref{31}) can then be rewritten as
\beqn
 S_{eff}^{aa}& =& \frac{1}{2\beta}\sum_{n=-\infty}^{\infty} \int \frac{dk_1}{2\pi} 
[ \tilde{\phi}_{n}^a(k_1) \tilde{\phi}_{-n}^a(-k_1)
           A_{n}^{aa}(k_1) \nonumber \\
           & + & \tilde{\eta}_{n}^a(k_1) \tilde{\eta}_{-n}^a(-k_1) 
           B_{n}^{aa}(k_1) + \tilde{\phi}_{n}^a(k_1)
           \tilde{\eta}_{-n}^a(-k_1) C_{n}^{aa}(k_1) ],
\label{40}
\eeqn 
where

 \beq
A_{n}^{aa}(k_1) = \lambda( \frac{\omega_{n}^2}{v_{F}} + v_{F} k_{1}^2) + 
                v_{F}^2 k_{1}^2~  \tilde{b}_{(0),n}^{aa}(k_1) +
           \tilde{b}_{(1),n}^{aa}(k_1) \frac{\omega_{n}^2}{v_{F}^2}
\label{41}
\eeq

\beq
B_{n}^{aa}(k_1) = \tilde{b}_{(0),n}^{aa}(k_1) \omega_{n}^2 + 
\tilde{b}_{(1),n}^{aa}(k_1) k_{1}^2
\label{42}
\eeq

\beq
C_{n}^{aa}(k_1) =2 [v_{F}\tilde{b}_{(0),n}^{aa}(k_1) - \frac{1}{v_{F}}
\tilde{b}_{(1),n}^{aa}(k_1)] \omega_{n}  k_1.
\label{43}
\eeq

The action (\ref{40}) can be easily diagonalized through the change 

\beq
\phi_{n}^a  =  \zeta_{n}^a - \frac{C_{-n}^{aa}}{2A_{-n}^{aa}}\xi_{n}^a \nonumber\\
\label{44}
\eeq

\beq
\eta_{n}^a  =  \xi_{n}^a
\label{45}
\eeq
which yields

\beqn
S_{eff}^{aa} & = & \frac{1}{2\beta} \sum_{n=-\infty}^{\infty} \int \frac{dk_1}{2\pi} 
 [\zeta_{n}^{a}(k_1) G_{\zeta,n}^{aa}(k_1) \zeta_{-n}^a(-k_1) + 
 \nonumber \\ 
 & + & \xi_{n}^a(k_1) G_{\xi,n}^{aa}(k_1) \xi_{-n}^a(-k_1)],
\label{46}
\eeqn
where

\beq
G_{\zeta,n}^{aa}(k_1)  =  \lambda ( v_{F} k_{1}^2 + \frac{\omega_{n}^2}{v_F})
+ [\tilde{b}_{(0)}^{aa}v_{F}^2 k_{1}^2 + \tilde{b}_{(1)}^{aa}\frac{ \omega_{n}^2}{v_{F}^2}]
\label{47}
\eeq
and
\beq
G_{\xi,n}^{aa}(k_1)  = \frac{ \lambda (\omega_{n}^2 / v_{F} 
+ v_{F}k_{1}^2) [\tilde{b}_{(0),n}^{aa} \omega_{n}^2  +
\tilde{b}_{(1),n}^{aa} k_{1}^2] + \tilde{b}_{(0),n}^{aa}\tilde{b}_{(1),n}^{aa} 
( v_{F}k_{1}^2 + \omega_{n}^2 / v_{F} )^2 } 
{ \lambda (v_{F}k_{1}^2 + \omega_{n}^2 / v_{F}) 
+ \tilde{b}_{(0),n}^{aa} v_{F}^2 k_{1}^2 + \tilde{b}_{(1),n}^{aa} 
\omega_{n}^2 / v_{F} }.
\label{48}
\eeq

The partition function of the system can now be expressed as
\beq
Z  =   N_{1}N_{F}(\beta) det_{\beta}^2(-\Box) det_{\beta}^2(
i~ \slp) \prod_{a=0,1} (detG^{aa}_{\zeta,n})^{-1/2} (detG^{aa}_{\xi,n})^{-1/2}
\label{49}
\eeq
and its logarithm, using the well known identity involving functional traces 
and determinants, can be written as

\beq
ln Z =  ln Z^{00} + lnZ^{11}
\label{50}
\eeq
where we have defined

\beqn
ln Z^{aa} & = &\frac{1}{2} ln \; N_{F}(\beta) + 
\frac{1}{2} ln \; N_{1} + tr \; ln(i {\slp}) +\nonumber\\
&+& tr \; ln({-\Box} ) - 
 \frac{1}{2}tr \;ln G_{\zeta,n}^{aa} - \frac{1}{2} tr \; ln G_{\xi,n}^{aa}
\label{51}
\eeqn
Since in general one is interested in derivatives of the partition functions 
with respect to temperature, we shall disregard $\beta$-independent 
terms. In particular, from now on any reference to $N_1$ will be omitted.

The last two terms in (\ref{51}) can be expressed in the form
\beqn
- \frac{1}{2}tr \;ln G_{\zeta,n}^{aa} - \frac{1}{2}
tr \; ln G_{\xi,n}^{aa} & = & - \frac{1}{2} tr \;ln [\tilde{b}_{(0),n}^{aa}(k_1) 
(\lambda + \frac{\tilde{b}_{(1),n}^{aa}(k_1)}{v_{F}})]
- \nonumber \\ & - & \frac{1}{2}tr \;ln [v_{F} k_{1}^2 + \frac{\omega_{n}^2}{v_{F}}] -
\nonumber\\
& - & \frac{1}{2} tr \; ln[\omega_{n}^2 + k_{1}^2 f_{n}^{aa}(k_1)]
\label{52}
\eeqn
with

\beq
f_{n}^{aa}(k_1) = \frac{{\tilde{b}}_{(1),n}^{aa}}{{\tilde{b}}_{(0),n}^{aa}} \;
\frac{ \lambda + v_{F}{\tilde{b}}_{(0),n}^{aa}}{ \lambda + 
{\tilde{b}}_{(1),n}^{aa}/v_{F}}
\label{53}
\eeq
Using also the well known identity \cite{ber}

\beqn
tr \; ln(v_{F} k_{1}^2 +\frac{1}{v_{F}} \omega_{n}^2) & = & 
\sum_{n=-\infty}^{\infty} \int \frac{dk_{1}}{2 \pi} ln (\frac{1}{v_{F}}
\omega_{n}^2 + v_{F} k_{1}^2) \nonumber \\
& = & tr \;ln ({-\Box}) =ln ~det_{\beta}(-\Box)
\label{54}
\eeqn
and replacing (\ref{52}) and (\ref{54}) in (\ref{51}) one obtains

\newpage

\beqn
ln Z^{aa} & = &\frac{1}{2} ln N_{F}(\beta) + ln \; det_{\beta}i {\slp} + 
\frac{1}{2} ln\; det_{\beta}({-\Box})  \nonumber \\ 
& - & \frac{1}{2} tr \; ln [{\tilde{b}}_{(0),n}^{aa}(k_1)( \lambda + 
\frac{{\tilde{b}}_{1}^{aa}}{v_{F}})] -  \nonumber \\
& - & \frac{1}{2} tr \; ln [\omega_{n}^2 
+k_{1}^2 f_{n}^{aa}(k_1)]
\label{55}
\eeqn

Let us recall that up to now we have considered a very general situation. 
Indeed, the above formula is valid for potentials that depend on both distances
and the temperature. However from a physical point of view it is reasonable to
assume that the interactions are temperature-independent. In this case we can 
generalize the procedure developed in Refs.\cite{ber} and \cite{dj} in order to write
\beqn 
\frac{-1}{2} tr~ln [\omega_{n}^2 + k_{1}^2 f^{aa}(k_1)] & = & \frac{1}{2}\int 
\frac{dk_1}{2\pi}\{ (1 -k \beta) (f^{aa})^{1/2}(k) - \nonumber \\
& - & 2 ln ( 1 - e^{-k \beta 
(f^{aa})^{1/2}(k)})  \nonumber \\ 
& + & 
2 ln ( 1 - e^{- (f^{aa})^{1/2}(k)}) \}  \nonumber \\
& - & \frac{1}{2} \int \frac{dk_1}{2 \pi} \sum_{n=-\infty}^{\infty} ln[(2 \pi n)^2 + 
f^{aa}(k)] \nonumber \\
& + & ln \beta \int \frac{dk_1}{2\pi} \sum_{n=-\infty}^{\infty},
\label{56}
\eeqn

\beqn
\frac{1}{2} ln \; det_{\beta} (-\Box) & = & \int \frac{dk_1}{2 \pi} 
\frac{v_{F} k \beta}{2} + \int \frac{dk_1}{2 \pi} ln ( 1 - e^{-v_{F} k \beta}) - 
\nonumber \\
& - & \int \frac{dk_1}{2 \pi} ln \beta\; \sum_{n=-\infty}^{\infty} 
\label{57}
\eeqn
and
\beqn
\frac{1}{2} ln \; det_{\beta} (i {\slp}) & = & \int \frac{dk_1}{2 \pi} 
\frac{v_{F} k \beta}{2} + \int \frac{dk_1}{2 \pi} ln ( 1 + e^{-v_{F} k \beta}) -  
\nonumber \\
& - & \int \frac{dk_1}{2 \pi} ln \beta\; \sum_{n=-\infty}^{\infty}  
\label{58}
\eeqn
where $k=\mid k_1 \mid$.
Please note that here we are calling $f^{aa}$ to the expression (\ref{53}) 
in the present, $n$-independent case. As usual, the $\beta$-dependent infinite 
contributions to the partition function are eliminated by choosing
\beq
ln N_{F}(\beta)=2 ln \beta \sum_{n=-\infty}^{\infty} \int \frac{dk_1}{2\pi}.
\label{59}
\eeq
Thus we finally obtain

\beqn
ln Z^{aa} & = &  \int \frac{dk_1}{2 \pi} \{\frac{3}{2} v_{F} k \beta + 
2 ln ( 1 + e^{-v_{F} k \beta}) + ln ( 1 - e^{-v_{F} k \beta}) \}
 + \nonumber \\ 
& + & \frac{1}{2} \int \frac{dk_1}{2 \pi} [(1- k \beta) (f^{aa})^{1/2}(k) -
2 ln ( 1 - e^{-k \beta (f^{aa})^{1/2}(k)}) + \nonumber \\
& + & 2 ln ( 1- e^{-(f^{aa})^{1/2}(k)})] - \frac{1}{2} \int \frac{dk_1}{2 \pi} 
\sum_{n=-\infty}^{\infty} ln [(2 \pi n)^2 + f^{aa}(k)]  - \nonumber \\
& - & \frac{1}{2} \int \frac{dk_1}{2 \pi} 
ln [\tilde b_{(0)}^{aa}(k) 
(\lambda + \frac{\tilde b_{(1)}^{aa}(k)}{v_{F}})] \sum_{n=-\infty}^{\infty} 
\label{60}
\eeqn

\noindent which is our first non-trivial result. Indeed, by inserting this 
expression in (\ref{50}) we have the partition function corresponding to the Thirring model 
with non-local interactions. Therefore, it is the extension of the well
known results for the local Thirring model at finite temperature 
(See \cite{rra}, \cite{mnt} and refs. therein) to the case in which non-contact 
potentials are taken into account. Apart from academic interest, this result 
could be useful in order to explore thermodynamical properties in strongly 
correlated systems, since for $b_{(1)}^{aa} = \infty$  the NLT model describes a TL 
system \cite{NLT}. 
Note that, in the case $b_{(0)}^{aa} = b_{(1)}^{aa} = b^{aa}(x,y)$ 
and $v_{F}=1$ 
(which leads to $f^{aa}(k)= 1$), 
one obtains a simplified 
non-local covariant model: 

\beqn
ln Z^{aa} & = & 2 \int \frac{dk_1}{2 \pi}\{ \frac{k \beta}{2} + ln ( 1 + e^{-k \beta})\}
- \nonumber \\
& - & \frac{1}{2} \int \frac{dk_1}{2 \pi} \sum_{n=-\infty}^{\infty} 
ln [\tilde b^{aa}(k)(\lambda + \tilde b^{aa}(k))]
\label{61}
\eeqn

\noindent which for $b_{0}^{aa}= b_{(1)}^{aa} = \delta^2(x-y)$ gives the right result
for the usual (local) Thirring model at finite temperature \cite{rra} 
\cite{mnt}:

\beqn 
ln Z_{Thirring} & =&
2 \int \frac{dk_1}{2 \pi} \{ \; \frac{k \beta}{2} +  \;ln ( 1 + e^{-k \beta}) - \nonumber \\
& -& \frac{1}{2} ln (\lambda + 1) \sum_{n=-\infty}^{\infty} \; \}
\label{62}    
\eeqn

\section{Thermodynamical functions}
\setcounter{equation}{0} 

In the previous Section we evaluated the partition function
for the non-local Thirring model. The main purpose of the present Section 
is to derive analytical expressions for relevant thermodynamical functions such as the 
Helmholtz free energy, the energy and the specific heat. Throughout this Section 
we shall consider $\beta$-independent potentials.
Let us begin with the Helmholtz free energy defined as  

\beq
F  =  - \frac{1}{\beta}lnZ  = F^{00} + F^{11}
\label{63}
\eeq
where
\beqn
F^{aa} & = & - \int \frac{dk_1}{2 \pi} \{ \frac{3k v_{F}}{2} + 
\frac{2}{\beta}ln ( 1 + e^{-k \beta v_{F}})  \nonumber \\ 
& + &  \frac{1}{\beta}ln (1-e^{-k \beta v_{F}}) - \frac{1}{\beta}
 ln ( 1 - e^{-k \beta (f^{(aa)})^{1/2}(k)})  \nonumber \\ 
& + &  \frac{(f^{aa})^{1/2}(k)}{2\beta}(1-k\beta) +\frac{1}{\beta} 
ln ( 1- e^{-(f^{aa})^{1/2}(k)}) \nonumber \\
& -  & \frac{1}{2\beta}  \sum_{n}
ln [(2 \pi n)^2 + f^{aa}(k)] - \nonumber\\ & - &
\frac{1}{2\beta}  ln [\tilde b_{(0)}^{aa}(k) 
(\lambda + \tilde b_{(1)}^{aa}(k))] \sum_{n=-\infty}^{\infty}  \}
\label{64}
\eeqn
The energy is given by $E=-\partial lnZ/\partial \beta$. Using
eq.(\ref{60}), ignoring the zero-point energy of the vacuum and taking into 
account that

\beq
2\int_{0}^{\infty} \frac{dk}{2\pi} \frac{k v_{F}}{e^{k\beta v_{F}}+1}=
 \int_{0}^{\infty} \frac{dk}{2\pi} \frac{k v_{F}}{e^{k\beta v_{F}}-1}=
 \frac{\pi}{12 \beta^2 v_{F}}
 \label{65}
 \eeq
we obtain

\beqn
E= E^{00} + E^{11} & = & \int \frac{dk_1}{2\pi} \{ \frac{(f^{00})^{1/2}(k) + 
(f^{11})^{1/2}}{2} +   \nonumber \\
& + & \frac{k (f^{00})^{1/2}(k)}{e^{k\beta (f^{00})^{1/2}}-1}  +
\frac{k (f^{11})^{1/2}(k)}{e^{k\beta (f^{11})^{1/2}}-1} \}
\label{66}
\eeqn
where, since we have supposed that the electron-electron potentials are 
$\beta$-independent, 
one has $f^{aa}(k,\omega)=f^{aa}(k)$.\\
Performing the derivative of the above expression with respect to temperature, 
one can write the 
specific heat as

\beqn 
C_V & = & \frac{1}{k_{B}T^2}\int \frac{dk_1}{2\pi} k^2 \{ f^{00}(k) 
\frac{e^{k (f^{00})^{1/2}(k)/k_{B}T}}{(1- e^{k (f^{00})^{1/2}(k)/k_{B}T})^2} + 
\nonumber \\
& + & f^{11}(k) \frac{ e^{k (f^{11})^{1/2}(k) /k_{B}T}}{(1- e^{k (f^{11})^{1/2}
(k)/k_{B}T})^2} \}
\label{67}
\eeqn

Recalling the dispersion relations for charge-density and spin-density modes given 
in \cite{NLT}, one can easily write $f^{00}$ and $f^{11}$ in terms of the 
corresponding velocities $v_{\rho}$ and $v_{\sigma}$ as

\beq
f^{00} = v_{F}^2 v_{\rho}^2
\label{68}
\eeq
and

\beq
f^{11} = v_{F}^2 v_{\sigma}^2.
\label{69}
\eeq

In order to check our general formula (\ref{68}) we can consider the local case, 
in which $v_{\rho}$ and $v_{\sigma}$ 
are constants. For this particular case we can easily compute the corresponding 
integral, which yields

\beq
C_{V} = k_{B}^2 \frac{\pi}{3} (\frac{1}{v_{\rho}} + \frac{1}{v_{\sigma}}) T
\label{70}
\eeq
which is the characteristic linear behavior of the 
underlying fermions (linear specific heat in any dimension) 
when only weak interactions (consistent with keeping only forward-scattering 
interactions) are taken into account.
Thus, in this Section we have presented analytical expressions for the energy and 
the specific heat of a general forward-scattering interacting system of fermions,
represented by the non-local Thirring model. Our formulae give the 
thermodynamical magnitudes not only as functions of temperature, but as 
functionals of the electron-electron potentials. Then, they could be used in 
order to examine the influence of 
different potentials on the equilibrium properties of the 1d electron liquid.

\section{Two-point fermionic correlations}
\setcounter{equation}{0}

\hs Let us now go back to the general case, i.e. with potentials not necessarily
$\beta$-independent, and compute the fermionic propagator defined by

\beq
<\Psi(\tau, x) \bP(\tau',y)>_{\beta} = \left( \begin{array}{cc}
                        0     &G_{+}^{\beta}(\tau,x,\tau',y) \nonumber \\
                  G_{-}^{\beta}(\tau,x,\tau',y) &   0
                  \end{array}   \right)
\label{71}
\eeq
where

\beq
G_{(\pm)}^{\beta}(\tau,x,\tau',y) = \left( \begin{array}{cc}
                         G_{(\pm)\uparrow}^{\beta}(\tau,x,\tau', y)   &0 \nonumber \\
                         0             &G_{(\pm)\downarrow}^{\beta}(\tau,x,\tau',y) 
                         \end{array}   \right) 
\label{72}
\eeq

The subindex $+(-)$ means that we consider right(left)-moving electrons, 
and $\uparrow (\downarrow)$ indicates that the field operator carries a spin 
up (down) quantum number. In the present case we have disregarded those 
processes with spin-flip. 
To be specific we shall restrict our analysis to $G_{(\pm) \uparrow}^{\beta}$ 
(similar expressions are obtained for $G_{(\pm) \downarrow}^{\beta}$).

 Once we have performed the decoupling change of variables given by eqs.
(\ref{25}) and (\ref{26})
the non-vanishing components of the thermal Green function are factorized 
in the form

\beqn
G_{\pm \uparrow}^{\beta}(x,\tau,y,\tau') =  <\Psi_{\pm \uparrow}(x) \bP_{\pm 
\uparrow}(y)>_{\beta} & = & G_{\pm \uparrow \beta}^{(0)}(x,\tau,y,\tau') 
B_{\pm \beta}^{00}(x,\tau,y,\tau') \times \nonumber \\
& \times & B_{\pm \beta}^{11}(x,\tau,y,\tau')
\label{73}
\eeqn
where $G_{ \pm \uparrow \beta}^{(0)}(x,\tau,y,\tau')$ is the free thermal Green 
function given by                                                                           

\beq
G_{\pm \uparrow \beta}^{(0)}(\tau-\tau',x-y) = e^{\pm iv_{F} p_{F}(x-y)} 
\sum_{n=-\infty}^{\infty}\frac{1}{2 \pi \beta}\int dk_{1} \frac {
e^{-i(\omega_{n}(\tau-\tau') + k_1 (x-y))}}{\omega_{n} \mp ik_1 v_{F}}
\label{74}
\eeq
with $\omega_{n} = (2n+1)\pi/\beta$, and

\beq
B_{\pm \beta}^{00} = < e^{g[\phi^{0}(y,\tau')-\phi^{0}(x,\tau)]} 
e^{ig[\eta^{0}(y,\tau')-\eta^{0}(x,\tau)]}>_{\pm \beta}^{00}
\label{75}
\eeq

\beq
B_{\pm \beta}^{11} = < e^{g[\phi^{1}(y,\tau')-\phi^{1}(x,\tau)]} 
e^{ig[\eta^{1}(y,\tau')-\eta^{1}(x,\tau)]}>_{\pm \beta}^{11}
\label{76}
\eeq

In equation (\ref{74}) we have redefined the energy origin, as usual, by 
introducing the Fermi momentum $p_F$. In equations (\ref{75}) and (\ref{76})
the symbol $<\;>^{aa}_{\beta}$ means v.e.v. with respect to the action 
(\ref{31}). Working in momentum space these bosonic factors can be written as

\beq
B_{\pm \beta}^{aa}(x,\tau,y,\tau') = \frac{\int D\tilde{\phi}^{a} 
D\tilde{\eta}^{a}~ e^{-[S_{eff}^{aa} 
+ S_{\pm \beta}^{aa}(x,\tau,y,\tau')]}}{\int D\tilde{\phi}^{a} D\tilde{\eta}^{a}~ 
e^{-S_{eff}^{aa}}} 
\label{77}
\eeq

\noindent with $S_{eff}$ given by eq.(\ref{40}) and

\beq
S_{\pm \beta}^{aa}(x,\tau,y,\tau ') = \frac{g}{\beta}\sum_{n} \int 
\frac{dk_1}{2\pi}~ 
[\pm \tilde{\phi}_{n}^{a}(k_1) + i \tilde{\eta}_{n}^{a}(k_1)] D
(\omega_{n},k_1;x,\tau,y,\tau ')
\label{78}
\eeq

\beq
D(\omega_{n},k_1;x,\tau,y,\tau ') = e^{i\omega_{n}\tau}e^{ik_{1}.x} - 
e^{i\omega_{n}(\tau ')}e^{ik_{1}.y}
\label{79}
\eeq

\noindent Now $B_{\pm \beta}^{aa}$ can be easily evaluated by performing the change

\beqn
\tilde{\phi}_{n}^{a}(k_1)& =& \tilde{\varphi}_{n}^{a}(k_{1}) + E_{n,\pm}^{aa}
(k_{1};x,\tau,y,\tau') \nonumber\\
\tilde{\eta}_{n}^{a}(k_1)& =& \tilde{\rho}_{n}^{a}(k_1) + 
F_{n,\pm}^{aa}(k_1;x,\tau,y,\tau')
\label{80}
\eeqn

\noindent where $\tilde{\varphi}^{a}$ and $\tilde{\rho}^{a}$ are the new quantum 
variables and $E_{n,\pm}^{aa}$ and $F_{n,\pm}^{aa}$ are classical 
functions chosen in the form

\beq
E_{n,\pm}^{aa}(k_1;x,\tau,y,\tau') = g[\frac{i C_{n}^{aa}(k_1)}{2} \mp 
B_{n}^{aa}(k_1)]\frac{D(\omega_{n},k_1;x,\tau,y,\tau')}
{\Delta_{n}^{aa}(k_1)}
\label{81}
\eeq

\beq
F_{n,\pm}^{aa}(k_1) = g[\pm \frac{C^{aa}(k_1)}{2} -iA_{n}^{aa}(k_1)]
\frac{D(\omega_{n},k_1;x,y)}{\Delta_{n}^{aa}(k_1)} 
\label{82}
\eeq

\noindent with $A_{n}^{aa}(k_1)$, $B_{n}^{aa}(k_1)$ and $C_{n}^{aa}(k_1)$ 
defined in ((\ref{41})-(\ref{43})) and

\beq
\Delta_{n}^{aa}(k_1) = (C_{n}^{aa})^2(k_1) - 4A_{n}^{aa}(k_1)B_{n}^{aa}(k_1)
\label{83}
\eeq

\noindent Putting all this together, the result is

\beqn
B_{\pm \beta}^{aa}(x,\tau,y,\tau') & = & exp\{-\frac{2g^2}{\beta} 
\sum_{n=-\infty}^{\infty} 
\int \frac{dk_1}{2 \pi}~ D(\omega_{n},k_1;x,\tau,y,\tau') \times \nonumber \\
& \times & D(\omega_{-n},-k_1;x,\tau,y,\tau') \times \nonumber\\
& \times & \frac{B_{n}^{aa}(k_1) - A_{n}^{aa}(k_1) \mp i C_{n}^{aa}(k_1)}
{\Delta_{n}^{aa}(k_1)}\}
\label{84}
\eeqn

\noindent which can be also written as

\beqn
B_{\pm \beta}^{aa}(x,\tau,y,\tau') & = & exp\{-\frac{8g^2}{\beta} 
\sum_{n=-\infty}^{\infty} \int \frac{dk_1}{2 \pi}~ 
sen^2 [\frac{\omega_{n}(\tau-\tau')}{2} + \frac{k_{1}(x-y)}{2}] \times\nonumber\\
& \times & \frac{B_n^{aa}(k_1) - A_n^{aa}(k_1) \mp i C_n^{aa}(k_1)}{\Delta_{n}^{aa}(k_1)}\}
\label{85}
\eeqn

This is the main result of this Section. Indeed, apart from the free
fermion contribution given by eq.(\ref{74}), eq.(\ref{85}) gives the fermionic 
thermal propagator as functional of the electron-electron potentials. Please 
note that up to this point these couplings are allowed to depend on both distance
and temperature.
Of course, in order to go further in this computation one needs to specify
the Fourier transformed inverse potentials $\tilde b_{(\mu),n}(k_1)$ which 
determine the integrand in (\ref{85}).
The simplest case at hand is the one corresponding to the usual 
Thirring model, which once again can be used to check the consistency of our more
general calculation. Setting $\tilde{b}_{(0)}^{aa} = \tilde{b}_{(1)}^{aa} = 1$, 
$v_{F} =1$, $X=(\tau,x)$, $Y=(\tau ',y)$ and $P=(\omega_{n},k_1)$
we get
\newpage
\beqn
B_{\pm \beta}^{Thirring}(x,\tau,y,\tau ') & = & exp\{-\frac{2\pi}{\beta}
\frac{(\frac{g^2}{\pi})^2}{(1+ \frac{g^2}{\pi})} \times \nonumber \\
& \times & \sum_{n=-\infty}^{\infty} \int \frac{dp}{2\pi} 
\frac{sen^2 \frac{P.(X-Y)}{2}}{P^2} \}
\label{86}
\eeqn

\noindent which, together with the corresponding free fermion contribution, 
can be easily shown to coincide with the well-known result for the local and 
covariant case \cite{rra} \cite{mnt}.

\section{Momentum distribution}
\setcounter{equation}{0}

 Once we have the fermionic two-point functions, it will be rather 
simple to get a closed expression for the electronic momentum distribution at 
finite temperature. Let us consider electrons with spin-up, whose distribution 
at finite temperature is given by

\beq
N_{\pm \uparrow}(q) = i \int^{+\infty}_{-\infty} dz_1~ e^{-iqz_1}
\lim_{z_0 \rightarrow 0} G_{\pm\uparrow}^{\beta}(z_0, z_1).
\label{87}
\eeq

where $Z=(z_0,z_1)=X-Y$.

Going back to (\ref{75}) we can take the limit $(\tau -\tau ')\rightarrow 0, 
(z_{0} \rightarrow 0) $ in $G_{\pm}^{(0) \beta}$ and perform the sum over 
$n$. We obtain

\beq 
lim_{z_{0} \rightarrow 0} G_{\pm \uparrow}^{(0) \beta}(z_{0},z_{1}) =
\frac{1}{4 \pi} e^{\pm iv_{F}p_{F}z_{1}} \int_{-\infty}^{+\infty} dk_1 
e^{-ik_1 z_{1}} cot (\frac{\pi}{2} \mp \frac{i \beta k_1 v_{F}}{2})
\label{88}
\eeq
Evaluating the $k_1$-integral we arrive at

\beq
\lim_{z_{0} \rightarrow 0} G_{\pm \uparrow}^{(0) \beta}(z_{0},z_{1}) =
\mp \frac{\pi}{2 \beta v_{F}} e^{\pm iv_{F}p_{F}z_{1}} cosech(\frac{\pi^2 z_{1}}{\beta v_{F}})
\label{89}
\eeq
Using this result together with the bosonic factors computed in the previous 
Section, the momentum distribution can be expressed as

\beqn
N_{\pm \uparrow}(q) & = & \mp \frac{i \pi}{2 \beta v_{F}} \int^{+\infty}_{-\infty} dz_1~ e^{-iz_1(q \mp v_{F}p_{F})}
cosech( \frac{\pi^2 z_{1}}{\beta v_{F}})  \nonumber \\
& \times & \exp{\left \{ \frac{-8}{\pi} \int dk_1 \; sin^2\frac{k_1 z_{1}}{2}
\left(I_{\pm}^{00}+I_{\pm}^{11}\right)(k_1)\right \}}
\label{90}
\eeqn
where we have defined $I_{\pm}^{ii} = \sum_{n= -\infty}^{\infty} 
\frac{[B_{n}^{ii} - A_{n}^{ii} \mp i C_{n}^{ii}]}{\Delta_{n}^{ii}}$.
At this point it is interesting to observe that for the physically relevant case 
in which the potentials are temperature-independent,
the sum can be readily evaluated, yielding:

\beqn
I_{\pm}^{ii}(k_1) & = & \frac{-\beta^2(v_{F}\lambda+\tilde b_{(1)}^{ii}v_{F}^{-2}
-\tilde b_{(0)}^{ii})}{16 \pi^2 \tilde b_{(0)}^{ii}(v_{F}\lambda + \tilde b_{(1)}^{ii})}
\frac{v_{F}^2\pi}{(T^{ii})^2 -S^2} \{ (S-R^{ii})cotg (\pi S) \nonumber \\
& - & (\frac{(T^{ii})^2 -R^{ii}S}{T^{ii}})cotg (\pi T^{ii})  \}
\label{91}
\eeqn    
where 

\beq
R^{ii}(k_1) = \frac{ i k_1 \beta v_{F}}{2 \pi}(\frac{ \lambda + v_{F}\tilde b_{(0)}^{ii} - \tilde b_{(1)}^{ii}v_{F}^{-1}}
{\lambda + \tilde b_{(1)}^{ii}v_{F}^{-1} - v_{F}\tilde b_{(0)}^{ii}}) \nonumber
\label{92}
\eeq

\beq
S(k_1) = \frac{i k_1 \beta v_{F}}{2 \pi} \nonumber
\label{93}
\eeq

\beq
T^{ii}(k_1) = \frac{i k_1 \beta}{2 \pi}[\frac{\tilde b_{(1)}^{ii}}
{\tilde b_{(0)}^{ii}}(\frac{v_{F}^{-1}\lambda 
+ \tilde b_{(0)}^{ii}}{v_{F}\lambda + \tilde b_{(1)}^{ii}})]^{1/2} \nonumber
\label{94}
\eeq

Thus, we finally get
\beqn
N_{\uparrow}(q) & = & N_{+\uparrow}(q) + N_{-\uparrow}(q) = \frac{\pi}{\beta v_{F}} 
\int_{-\infty}^{+\infty}
dz_{1} e^{-i q z_{1}} sin (p_{F}z_{1})  \times \nonumber \\ 
& \times & cosech (\frac{\pi^{2} z_{1}}{\beta v_{F}})
\exp\left \{\frac{-8}{\pi}\int dk_1 \;
sin^2 \frac{k_1 z_{1}}{2}(I_{+}^{00}+I_{+}^{11})(k_1)\right \}
\label{95}
\eeqn

Of course, in order to have the complete result one has to add the contribution
corresponding to spin-down electrons. In the present situation 
(no spin-flipping interactions) it is easy to see that both up and down results coincide
and therefore, (\ref{95}) must be multiplied by $2$.\\
This result could be used to perform quantitative studies of the momentum 
distribution in the context of 1d 
many-body systems \cite{gut}, as functional of potentials. In fact, this 
formula is the extension to the $T\neq{0}$ case of the one first presented in
\cite{NLT}. As such, it could be helpful in order to verify if the interplay between 
electron-electron interactions 
and thermal corrections could give rise to a
restoration of Fermi liquid behavior, as suggested in the literature \cite{das sarma},\cite{imp}.

\section{Conclusions}

In this work we have extended a recently proposed approach to the bosonization
of a Thirring model with non-local current-current interaction, originally 
developed at $T=0$ \cite{NLT}, to the finite-temperature case. In view of the 
connection between this model and a one-dimensional system of fermions (in fact 
our non-local model coincides with the Tomonaga-Luttinger model for a particular 
choice of the coupling potentials), the present study is relevant not only from a
purely academic point of view but also as a possible starting point to examine
thermodynamical properties of a Luttinger liquid through an alternative 
field-theoretical formulation.  

In Section 2 we have shown how the standard techniques of finite temperature 
local and covariant QFT's \cite{ber} \cite{mat} \cite{dj} can be also 
succesfully implemented in the non-local Thirring model. We obtained an 
expression for the partition function as functional of the forward-scattering
potentials. From this result, in Section 3 we derived the corresponding formulae
for the Helmholtz free energy, the energy and the specific heat. In Sections 4 
and 5 we computed the two-point fermionic correlation and the momentum 
distribution, respectively.

One interesting aspect of our results is that they provide practical formulae to
check the validity of different potentials. Moreover, they could be used to 
explore the interplay between thermal effects and electron-electron scattering 
(through suitable potentials). However, in order to make these studies more
realistic one should include two ingredients that were disregarded in this 
paper: spin-flippings and backward-scattering. We hope to report on some of these
issues in the close future.

\section{Acknowledgements}
The authors are financially supported by Universidad Nacional de La Plata (UNLP) 
and Consejo Nacional de Investigaciones Cient\'{\i}ficas y T\'ecnicas\\
(CONICET), Argentina.


\newpage

\begin{thebibliography}{99}
\bibitem{Voit} J. Voit, Rep. Prog.Phys. {\bf58}, 977 (1995).
\bibitem{Haldane} F. Haldane, J. Phys. C{\bf14}, 2585 (1981).
\bibitem{TL} Tomonaga, Prog.Theor.Phys.{\bf5}, 544 (1950).\\
J.Luttinger, J. Math. Phys. {\bf4}, 1154 (1963).\\  
E. Lieb and D. Mattis, J. Math. Phys. {\bf 6} (1965) 304. 
\bibitem{NLT} C.M.Na\'on, M.C.von Reichenbach and M.L.Trobo,
 Nucl. Phys.[FS]{\bf B435}, 567 (1995).\\
D.G.Barci and C.M.Na\'on, 
Int. J. Mod. Phys. A, in press (1997),(hep-th/9705077).
\bibitem{ber} C.Bernard, Phys.Rev.D{\bf9}, 3312 (1974).
\bibitem{mat} T.Matsubara, Prog.Theor.Phys.{\bf14}, 351 (1955).
\bibitem{mnt} M.V.Man\'{\i}as, C.M.Na\'on and M.L.Trobo, Phys.Lett.B (1997),
in press (hep-th 9701160), and references therein. 
\bibitem{Klaiber} B.Klaiber, Lectures in Theoretical Physics, Boulder, Colorado,
1967, Vol.10A, eds. A.Barut and W.Brittin (Gordon and Breach, New York), p.141.
\bibitem{So} J.S\'olyom, Adv. in Phys.{\bf28}, 201 (1979).
\bibitem{rd} M.Reuter and W.Dittrich, Phys.Rev.D{\bf32}, 513 (1985).
\bibitem{dj} L.Dolan and R.Jackiw, Phys.Rev.D.{\bf9}, 3320 (1974).
\bibitem{rra} F.Ruiz-Ruiz and R.F.Alvarez-Estrada, Phys.Rev.D{\bf35},
3161 (1987).
\bibitem{gut} H.Gutfreund and M.Schick, Physical Review {\bf168}, 418
(1968).
\bibitem{das sarma} B.Hu and S. Das Sarma, Phys.Rev.B{\bf48}, 5469 (1993).
\bibitem{imp} C.M.Na\'on, M.C.von Reichenbach and M.L.Trobo,
Nucl. Phys.[FS] {\bf B485}, 665 (1997).

\end{thebibliography}
\end{document}